\documentclass[journal]{IEEEtran}

\usepackage{cite}
\usepackage[T1]{fontenc}
\usepackage{amsmath,amssymb,amsfonts}
\usepackage{algorithmic}
\usepackage{graphicx}
\usepackage{textcomp}
\usepackage{multirow}
\usepackage{tabularx}
\usepackage{tabu}                      
\usepackage{booktabs}                  
\usepackage{makecell}
\usepackage{hyperref}

\usepackage{tikz}
\usepackage{tikzscale}
\usetikzlibrary{arrows,external}
\usetikzlibrary{arrows.meta}
\usetikzlibrary{positioning}
\usetikzlibrary{spy}
\usetikzlibrary{calc}
\usetikzlibrary{decorations,calligraphy}
\usepackage[capitalize]{cleveref}
\crefname{section}{Sec.}{Secs.}
\Crefname{section}{Section}{Sections}
\Crefname{table}{Table}{Tables}
\crefname{table}{Tab.}{Tabs.}

\newcommand{\tileFraction}{0.1}
\newcommand{\tileFractionCom}{0.1}
\newcommand{\tileFractionAb}{0.1}
\newcommand{\archFraction}{1.0}

\begin{document}
\title{Structurally Consistent MRI Colorization using Cross-modal Fusion Learning}

\author{\textit{Mayuri Mathur$^1$, Anav Chaudhary$^2$, Saurabh Kumar  Gupta$^3$, and Ojaswa Sharma$^1$} \\$^1$Indraprastha Institute of Information Technology Delhi, India \\$^2$Purdue University, USA \\$^3$All India Institute of Medical Sciences, India }
\maketitle

\maketitle 

\begin{abstract}
  Medical image colorization can greatly enhance the interpretability of the underlying imaging modality and provide insights into human anatomy. The objective of medical image colorization is to transfer a diverse spectrum of colors distributed across human anatomy from Cryosection data to source MRI data while retaining the structures of the MRI. To achieve this, we propose a novel architecture for structurally consistent color transfer to the source MRI data. Our architecture fuses segmentation semantics of Cryosection images for stable contextual colorization of various organs in MRI images. For colorization, we neither require precise registration between MRI and Cryosection images, nor segmentation of MRI images. Additionally, our architecture incorporates a feature compression-and-activation mechanism to capture organ-level global information and suppress noise, enabling the distinction of organ-specific data in MRI scans for more accurate and realistic organ-specific colorization. Our experiments demonstrate that our architecture surpasses the existing methods and yields better quantitative and qualitative results. Our source code will be released here \href{https://github.com/graphics-research-group/MRI-Colorization-Cross-Modal-Fusion-Learning.git}{https://github.com/graphics-research-group/MRI-Colorization-Cross-Modal-Fusion-Learning.git}
 
\end{abstract}

\section{Introduction}
\label{sec:introduction}

Radiometric images like MRI and CT contain grayscale pictures of human anatomy used for non-invasive clinical diagnosis. Only trained radiologists can visually recognize and distinguish anatomical structures in radiometric imaging data. Recently, deep-learning techniques have offered automated organ-based structural recognition for color transfer in the medical domain and improved the visual experience of human anatomy. Extensive work has been done for natural image colorization by utilizing deep-learning techniques \cite{vitoria2020chromagan,ji2022colorformer,Kang2023ICCV}, however these techniques underperform in the medical domain. The existing colorization architectures are more suited for natural images and their performance degrades when adapted to the medical domain. This is primarily due to the fact that many such approaches \cite{ji2022colorformer,Kang2023ICCV} incorporate pre-trained networks like VGG \cite{simonyan2014very} that are trained only on natural images. There are limited contributions to medical image colorization. The existing medical colorization methods have limited color variations in anatomical structures, some techniques \cite{Mathur2021WACV} use 3D convolutional layers in their frameworks with larger memory requirements for optimization. Some techniques \cite{wang2022colorizing,haq2022novel} colorize partial sections of the body efficiently but are incapable of the diverse color synthesis for anatomical structures of the full body. 
While most methods \cite{chen2019synergistic,dong2019synthetic,huo2018synseg, ganeshkumar2022identification} focus on segmentation which cannot be used for colorization since
these methods only capture global information and are inefficient in fetching diverse colors and textures of the various organs. 
We propose a novel semi-supervised architecture for structurally consistent MRI colorization to overcome the aforementioned limitations in the medical domain. In our method, the MRI data is not completely registered with the Cryosection data. Deformations persist in the MRI data even after performing a rigid registration followed by deformable registration using B-splines deformation, therefore, the available Cryosection segmentation data cannot be paired with MRI data.
Our method builds the cross-domain organ-color and organ-structure association between partly registered heterogeneous data. It integrates the Cryosection segmentation semantics with the perceptually similar semantics of the MRI  data during cyclic generation across domains. This is achieved by the color adaptation dual decoder module used in our proposed colorization generator in the forward pass of our cycle-consistent network for bridging multimodal information of colors and structures. One decoder is designed for building an association between organ regions and colors using the semantics of segmentation data while the other decoder performs MRI colorization by leveraging this association during optimization.
The source MRI data contains noise that introduces ambiguity in differentiating between various organ textures that disrupt the colorization process. To overcome this limitation and to further enhance the performance of building color and anatomical region association in our architecture, we introduce skip connections between the two decoders using a compression-activation mechanism that suppresses noise and captures the global information of diverse organs. To the best of our knowledge, our architecture provides full body colorization of partly registered MRI data. The major contributions of our work are as follows:
\begin{itemize}
    \item A novel architecture to colorize the whole body that can effectively capture anatomical structures of the MRI data and colors of Cryosection data.
    \item  Integration of Cryosection segmentation information in the proposed cross-modality adaptation framework for a robust semantic and color-texture correlation.
    \item A multiscale module to handle varying resolutions of input MRI images.
\end{itemize}


\section{Related work}
\paragraph{Cross modality synthesis} Various deep-learning-based techniques \cite{8759529,hiasa2018cross,9629705,8902799,Kim2024WACV,chen2019synergistic,9761546} are utilized in cross-modality adaptation frameworks in the medical domain. Ge et al. \cite{8759529} perform unpaired PET-MR image to CT image synthesis by enforcing structural constraints by incorporating Mutual information (MI) with CycleGAN. Hiasa et al. \cite{hiasa2018cross} performed cross-modality synthesis on MRI and CT data using CycleGAN with gradient loss as a similarity metric to impose structural alignment for multimodal registration. Felfeliyan et al. \cite{9629705} demonstrate translating the source domain (COR IW 2D TSE Fat Suppressed MRI) to the target domain (Sagittal 3D DESS WE) via CycleGAN for unsupervised segmentation of the source domain. It incorporates an instance segmentation network (I-MaskRCNN). Cycle-MedGAN \cite{8902799} extends CycleGAN by implementing perceptual loss for reducing semantic disparities between the input and the cycle reconstructed output and cycle style loss for fine detail enhancement in the cycle reconstructed output. Kim and Park \cite{Kim2024WACV} perform multimodal domain adaptation of MRI data using a 3D latent diffusion model which uses a multiple switchable spatially adaptive normalization (MS-SPADE) block for modality translation. Synergistic Image and Feature Adaptation (SIFA) \cite{chen2019synergistic} leverages cyclic modality fusion for medical segmentation by fusing the image and feature spaces. It requires semantic segmentation information from both modalities for cross-domain fusion. APS~\cite{9761546} is a cross-modality adaptation technique used for MRI-to-CT image synthesis. This method uses adversarial loss, pixel translation loss, and structural consistency loss during optimization.  

\paragraph{Colorization}
Several colorization algorithms exist for colorizing natural data and have limited contribution to the automation of colorization in the medical domain. 
Deep learning-based colorization methods \cite{wang2022colorizing,Mathur2021WACV,Carrillo2022ACCV,vitoria2020chromagan,ji2022colorformer,Kang2023ICCV} involve training a machine learning model, such as a Convolutional Neural Network (CNN), to automate color transfer from reference color image to grayscale image. Wang et al. \cite{wang2022colorizing} proposed a learning-based colorization method that aims to colorize grayscale lung CT images using  ResNet architecture. Mathur et al. \cite{Mathur2021WACV} proposed a framework for exemplar-based 2D to 3D colorization of MRI image via grayscale modality conversion, style transfer, and optimization-based colored volume synthesis. Carrillo et al. \cite{Carrillo2022ACCV} automate colorization using super attention blocks \cite{carrillo2022non} in U-Net architecture by using a color reference image as an exemplar. ChromaGAN~\cite{vitoria2020chromagan} is an unsupervised learning technique that adapts class distribution information for the colorization of natural images which utilizes Wasserstein Generative Adversarial loss (WGAN), semantic-color loss using Euclidean norm, and   Kullback-Leibler divergence for class distribution learning. ColorFormer~\cite{ji2022colorformer} incorporates a transformed-based architecture to capture semantic context for the colorization of natural images. The technique involves using a memory decoder to obtain a diverse color spectrum for colorization. The memory decoder stores multiple groups of color priors for color diversity as key-value pairs. DDColor~\cite{Kang2023ICCV} is another transformer-based colorization technique for natural images that uses dual decoders. One decoder restores semantic information from a feature pyramid obtained by stacking features from the upsampling layers. The other decoder improves the color correspondence with the semantic details.
\paragraph{Data compression techniques}
Squeeze-and-excitation (SE) networks \cite{hu2018squeeze} fine-tune the features from neural networks to capture the global distribution of the input to the network for performing segmentation, classification, and object detection.
SE networks \cite{zhang2020classification,roy2018recalibrating,chen2023dusfe,roy2018recalibrating,li2020deepseed,xiong2024sea} have proven to enhance the efficiency of image segmentation and image classification upon deploying them with existing deep learning architectures in the medical imaging domain. 

Zhang et al. \cite{zhang2020classification} perform classification on lung nodules for identifying the malignant and benign nodules by deploying SE blocks in the classification architecture ResNeXt as SE-ResNeXt network.  Roy et al. \cite{roy2018recalibrating} perform 2D MRI segmentation by incorporating channel squeeze and spatial excite mechanism. The authors propose spatial attention technique,  spatial and channel SE (scSE) blocks which reconfigure the feature to map to both channel and spatial spaces.  DeepSEED \cite{li2020deepseed} is a deep convolutional neural network for pulmonary nodule detection that utilizes SE-based encoder-decoder model. It is based on the ResNet architecture and incorporates SE mechanism for each convolutional layer of the architecture. SEA-Net \cite{xiong2024sea} performs medical image segmentation for different modalities using SE techniques in their deep neural network. It incorporates an attention path and an SE path to merge the deep semantic and the shallow semantic information of the input. The attention path is positioned above every SE block incorporated in skip connections of the UNet architecture. 

\section{Approach}

We perform experiments on a subset of the Visible Korean Human dataset \cite{VKHanothertrial,VKHIEEE,VKHTechniques} that contains Cryosection, CT, MRI, and Cryosection segmentation images of a male subject. For our approach, we perform MRI volumetric registration with Cryosection volume by using deformable B-spline registration available in the Elastix \cite{klein2009elastix} library. The Cryosection segmentation data consists of 54 subclasses corresponding to 13 anatomical systems of the human body.

Our method aims to perform colorization by transferring textural details of the Cryosection image to an MRI image with minimum structural alterations in the latter. It comprises of generative adversarial network that leverages cyclic cross-modality fusion technique \cite{zhu2017unpaired}. Our architecture offers a dual decoder module in the color-transferring generator to adapt distinguishable colors and textures of various organs. It also facilitates cross-modal correlation between the segmentation of Cryosection data and perceptually similar structures of partly registered data.  We design skip connections of our architecture with compression and excitation mechanism \cite{hu2018squeeze} to capture global semantic information about organ level and noise suppression. We extend the capability of our architecture for multiscale invariance to interpret organ details at different resolutions by incorporating the hierarchical spatial variation module in the encoder of the colorization generator.

\begin{figure*}[!htp]
    \centering
    \includegraphics[width= \archFraction\linewidth]{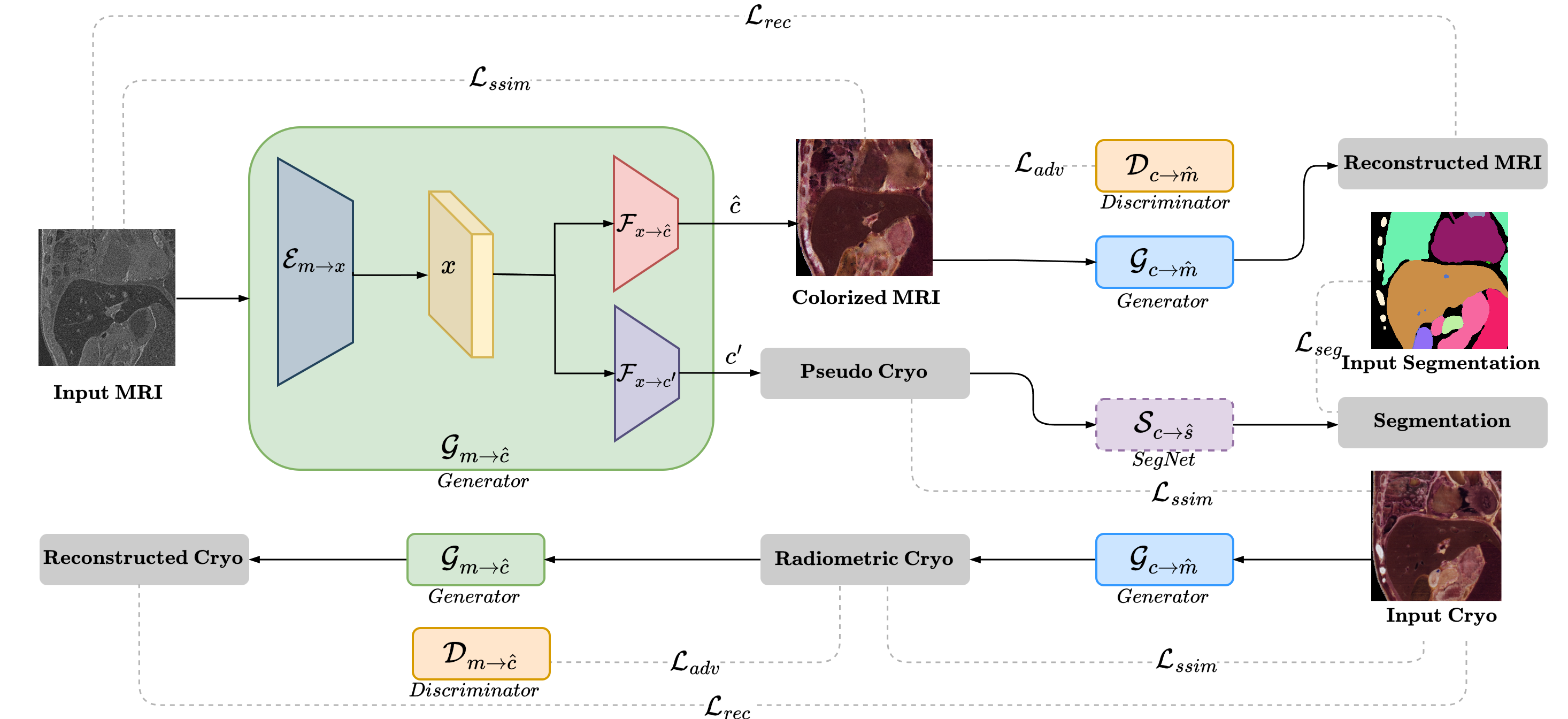}
    
    \caption{Architecture of our cycle consistent MRI colorization training model. The primary generator module $\mathcal{G}_{m\rightarrow \hat{c}}$ produces colorized MRI and an auxiliary output of pseudo Cryosection image which is used to build organ level context in the network. We use a pre-trained segmentation network $\mathcal{S}_{c\rightarrow \hat{s}}$. Various losses are indicated with light dashed lines.}
    \label{fig:overview}
\end{figure*}

\subsection{Colorization architecture}
Our proposed colorization architecture is designed using the generative adversarial network with a cyclic cross-modality adaptation technique \cite{zhu2017unpaired}, as shown in as shown in Fig.~\ref{fig:overview}. Our architecture offers a colorization generator comprising two pathways by incorporating dual decoders. The two pathways of the cyclic cross-modality translation architecture are MRI to colorized MRI synthesis and Cryosection to radiometric Cryosection synthesis. The first pathway performs color transfer of ground truth Cryosection data to input MRI and the second pathway transfers the grayscale properties of the input MRI to input Cryosection data. Our method modifies the generator operating on MRI to colorized MRI synthesis pathway by incorporating two decoders into this generator instead of the conventional single decoder and integrates the segmentation information with this generator. This improves the performance of mapping the structural and color information of MRI and Cryosection data, respectively.  Let us denote the training dataset images of MRI, Cryosection, and segmentation by $M, C$, and $S$ respectively. The cyclic consistency technique comprises two generators: $\mathcal{G}_{m\rightarrow \hat{c}}$ and $\mathcal{G}_{c\rightarrow \hat{m}}$ operating on Cryosection image $c$ sampled from its data distribution $p_c$, and MRI image $m$ sampled from its data distribution $p_m$ respectively. The first generator $\mathcal{G}_{m\rightarrow \hat{c}}: M\mapsto C$ oversees the colorization of MRI domain by adopting color distribution of the Cryosection detailed organ-specific color synthesis. In addition to a colorized MRI, it outputs an auxiliary pseudo Cryosection image. The second generator model $\mathcal{G}_{c\rightarrow \hat{m}}: C\mapsto M$ supervises the mapping of intensities from the MRI domain to the Cryosection domain. The pseudo Cryosection synthesis is used for correlating the semantics of the partly registered input MRI and ground truth Cryosection to leverage the segmentation information only available for the ground truth Cryosection for the organ-level structural and color association. During training, the weights of both decoders are updated using cross-domain information to establish the semantic association. Our approach modifies one of the cross-modality generators, $\mathcal{G}_{m\rightarrow \hat{c}}$ that guides the colorization pathway of the entire cyclic architecture. For this generator, the encoder $\mathcal{E}_{m\rightarrow x}$ is a mapping that converts an input MRI image $m$ to a bottleneck feature $x$ that is forwarded as common input to two decoders: $\mathcal{F}_{x\rightarrow \hat{c}}$ and $\mathcal{F}_{x\rightarrow c^\prime}$. The colorized MRI $\hat{c}$ is structurally similar to input MRI $m$ and colors are similar to $c$ while both colors and structures of pseudo Cryosection $c^\prime$ are similar to $c$. We input $c^\prime$ to our pre-trained Cryosection segmentation network based on U-Net \cite{ronneberger2015u} architecture $\mathcal{S}_{c\rightarrow \hat{s}}: C\mapsto S$ which is trained on the pairs of images $(C, S)$ where $\hat{s}$ is the label map extracted for input $c$. The parameters of $\mathcal{S}_{c\rightarrow \hat{s}}$ remained fixed during optimization of the cyclic consistency models. This aids class distribution learning to enhance color and texture mapping of the Cryosection anatomical structures with the approximated segmentation of the colorized MRI.

\subsection{Compression-activation generators} 
Our proposed framework implements channel compression and activation mechanisms  \cite{hu2018squeeze} to all the skip connections of its architecture. The compression-activation blocks incorporate linear layers that replace the conventional convolutional residual blocks. The compression-activation technique is used to capture the organ-level global information. As this technique emphasizes the important features, it tends to suppress less useful ones. Here, it reduces the impact of noise in the MRI data. The ability to capture global information of anatomical classes is leveraged to maintain the structural consistency of the source MRI data in the colorized MRI data. In this mechanism, compression extracts channel-wise descriptor embedding by aggregating the spatial and channel-wise information of the feature from the preceding convolutional layer. Activation is performed by applying a sigmoid-based gating mechanism to the channel-wise descriptor embedding. The sigmoid-activated embedding is used to reweigh the input convolutional feature, which is passed as input to the next convolutional layer. 

\subsection{Multiscale module}
We implement a multiscale module in the generators $\mathcal{G}_{m\rightarrow \hat{c}}$ and $\mathcal{G}_{c\rightarrow \hat{m}}$, and pass three inputs corresponding to the original resolution and its downsampled versions. Both the input MRI image $m$ and the input Cryosection image $c$ are downsampled by a factor of two resulting in $m_2\in\mathbb{R}^{\frac{w}{2}\times \frac{h}{2}}$ and $m_4\in\mathbb{R}^{\frac{w}{4}\times \frac{h}{4}}$ where  $w$ and $h$ correspond to height and width of $m$ respectively and the same is implemented for the input Cryosection $c$ keeping the number of channels intact. The multiscale module comprises three convolutional submodules where each submodule operates on each $m$, $m_2$ and $m_4$ respectively. The features extracted from these submodules are fused by summing the feature maps of $m_2$ and $m_4$ and concatenating with the feature of $m$. The fused feature is passed to the succeeding layers of the generator.   

\subsection{Approach formulation}
Our approach focuses on colorizing MRI images while preserving structural integrity across domain shifts. We accomplish this by optimizing the fidelity of our colorization using cyclically consistent adversarial loss, structural similarity loss, and segmentation loss.

\paragraph{Cyclic consistency loss} 
We incorporate cyclic cross-modality adaptation losses \cite{zhu2017unpaired} which are adversarial loss $\mathcal{L}_{adv}$ and cycle consistency loss $\mathcal{L}_{rec}$. Modality translation is achieved by minimizing the adversarial loss \cite{goodfellow2014generative} within the mappings $\mathcal{G}_{m\rightarrow \hat{c}}$ and $\mathcal{G}_{c\rightarrow \hat{m}}$
\begin{align}
   &\mathcal{L}_{adv} = \nonumber\\
   &\mathbb{E}_{c\sim p_c}\left[\log \mathcal{D}_{c}(c)\right] +\mathbb{E}_{m\sim p_m}\left[\log(1 - \mathcal{D}_{c}(\mathcal{G}_{m\rightarrow \hat{c}}(m)))\right] + \nonumber\\
   &\mathbb{E}_{c\sim p_c}\left[\log \mathcal{D}_{m}(m)\right] +\mathbb{E}_{m\sim p_m}\left[\log(1 - \mathcal{D}_{m}(\mathcal{G}_{c\rightarrow \hat{m}}(c)))\right],
\end{align}
where $\mathcal{D}_c$ and $\mathcal{D}_m$ discriminate the cross-modal distribution adaptability by classifying the synthesized data of respective data as real and pseudo. Reconstruction loss used in cycle consistency is computed between the input and the reconstructed modalities. For an MRI sample $m$, a cascade of generators produce the reconstructed MRI image $\hat{m} = \mathcal{G}_{c\rightarrow \hat{m}}(\mathcal{G}_{m\rightarrow \hat{c}}(m))$. Similarly, for a Cryosection image $c$, the reconstructed Cryosection image $\hat{c} = \mathcal{G}_{m\rightarrow \hat{c}}(\mathcal{G}_{c\rightarrow \hat{m}}(c))$. 
We formulate the reconstruction loss as
\begin{align}
    \mathcal{L}_{rec} = &\mathbb{E}_{c\sim p_c}\left[ \left\| \mathcal{G}_{m\rightarrow \hat{c}}(\mathcal{G}_{c\rightarrow \hat{m}}(c)) - c \right\|_1\right] +\nonumber\\
    &\mathbb{E}_{m\sim p_m}\left[ \left\| \mathcal{G}_{c\rightarrow \hat{m}}(\mathcal{G}_{m\rightarrow \hat{c}}(m)) - m \right\|_1\right].
\end{align}
The total Cyclic Adversarial Loss is then defined as 
\begin{equation}
    \mathcal{L}_{cyc} = \mathcal{L}_{adv} + \mathcal{L}_{rec},
\end{equation}

\paragraph{Structural Adaptation Loss} 
Our method incorporates a variant of Structural Similarity Index Measure (SSIM) \cite{wang2004image} for structurally consistent color adaptation. We compute a local SSIM map using an image's luminance, contrast, and structure over small patches of variable sizes. In our approach, we compute the local SSIM values over four patch sizes of $3\times 3$, $5\times 5$, $7\times 7$, and $9\times 9$ and aggregate these four resulting SSIM maps in the loss term. The local SSIM value at an image index $(i, j)$ over a patch is defined as follows
\begin{align}
    \textrm{SSIM}_{ij}(m, \hat{c})= l_{ij}\left ( m, \hat{c} \right )k_{ij}\left(m, \hat{c} \right )t_{ij}\left ( m, \hat{c}\right),
\end{align}

where $l_{ij}\left ( m, \hat{c} \right )$ denotes the luminance similarity, $k_{ij}\left ( m, \hat{c} \right )$ denotes the contrast similarity and $t_{ij}\left ( m, \hat{c} \right )$ denotes the local structural similarity between $m$ and $\hat{c}$ over a patch:
\begin{align}
    l_{ij}\left ( m, \hat{c} \right ) &= \frac{2\mu_{m, ij}\mu_{\hat{c}, ij}+C_{1}}{\mu_{m, ij}^{2} +\mu_{\hat{c}, ij}^{2}+C_{1}},\nonumber\\
    k_{ij}\left ( m, \hat{c} \right ) &=\frac{2\sigma_{m, ij}\sigma_{\hat{c}, ij}+C_{2}}{\sigma_{m, ij}^{2}+\sigma_{\hat{c}, ij}^{2}+C_{2}},\nonumber\\
    t_{ij}\left(m,\hat{c}\right) &=\frac{\sigma_{m\hat{c}, ij}+C_{3}}{\sigma_{m, ij}\sigma_{\hat{c}, ij}+C_{3}}.\nonumber
\end{align}
Here, $\mu_{m, ij}$ and $\mu_{\hat{c}, ij}$ represent the mean values, $\sigma_{m, ij}$ and $\sigma_{\hat{c}, ij}$ represent the standard deviations, and $\sigma_{m\hat{c}, ij}$ represents the covariance between $m$ and $\hat{c}$ over a patch centered at image index $(i, j)$. Constants $C_{1}, C_{2}$, and $C_{3}$ are small positive constants to prevent instability when the denominator approaches zero (see Wang et al.~\cite{wang2004image} for details).

We compute three SSIM losses $\mathcal{L}_{ssim}(m, \hat{c})$, $\mathcal{L}_{ssim}(c, \hat{m})$, and $\mathcal{L}_{ssim}(c, c^\prime)$ for enforcing structural similarity between the pairs of images ($m$, $\hat{c}$), ($c$, $\hat{m}$), and ($c$, $c^\prime$), respectively. $\mathcal{L}_{ssim}(m, \hat{c})$ is defined as the mean of all values in a local SSIM map summed over different patch sizes:
\begin{align}
\mathcal{L}_{ssim}(m, \hat{c}) = \sum_{b\in\{3, 5, 7, 9\}}\mathrm{mean}(\textrm{SSIM}_{b\times b}(m, \hat{c})).
\end{align}
$\mathcal{L}_{ssim}(m, \hat{c})$ and $\mathcal{L}_{ssim}(c, c^\prime)$ are computed similarly. The total SSIM loss is given by
\begin{align}
    \mathcal{L}_{ssim} = \underbrace{\mathcal{L}_{ssim}(c, \hat{m})}_{\text{from $G_{c\rightarrow \hat{m}}$}} + \underbrace{\mathcal{L}_{ssim}(m, \hat{c}) +  \mathcal{L}_{ssim}(c, c^\prime)}_{\text{from $G_{m\rightarrow \hat{c}}$}}.
\end{align}

\paragraph{Segmentation loss}
The Visible Korean Human dataset provides segmentation data corresponding to the Cryosection images. The same cannot be paired with the MRI data due to non-rigid deformations between Cryosection and MRI images. We unify the feature space of both MRI and segmentation data via a pseudo Cryosection image $c^\prime$ synthesized from the decoder 
 $\mathcal{F}_{x\rightarrow c^\prime}$. The  cross-entropy (CE) loss function  \cite{ronneberger2015u} $\mathcal{L}_{seg}$ used to update parameters of $\mathcal{G}_{m\rightarrow \hat{c}}$ is given by
\begin{align}
    \mathcal{L}_{seg} = \textrm{CE}(s, \mathcal{S}_{c\rightarrow \hat{s}}(c)),
\end{align}
where CE between two multi-class binary maps $s, \hat{s} \in \mathbb{R}^3\mapsto\{0,1\}$ is defined as:
\begin{align}
    \textrm{CE}(s, \hat{s}) = -\frac{1}{wh} \sum_{k}^{l}\sum_{j}^{h}\sum_{i}^{w}s_{ijk}\log \hat{s}_{ijk},
\end{align}
where $l$ is the total number of segmentation classes. The complete objective of our approach is given by loss $\mathcal{L}~=~\mathcal{L}_{cyc} + \mathcal{L}_{ssim} + \mathcal{L}_{seg}$.

\section{Results and analysis}
Our model consists of convolutional layers along with residual connections. The discriminator consists of five convolutional layers with LeakyReLu activations followed by a final linear layer with  Sigmoid activation. The input spatial size of the multimodal networks is $256\times 256$. The segmentation dataset has 46 accurate segmentations of Cryosection images. We optimize all models using the Adam optimizer, with learning rates set to $10^{-3}$, $10^{-4}$, and $10^{-6}$ for generators, discriminators, and segmentation models respectively. Training is conducted for 150 epochs with a batch size of 28  using Pytorch with distributed learning across two Nvidia A100 GPUs with 80 GB of memory each. We extract $12$,$000$ slices of resolution $(256 \times 256)$ at random planes from the volumetric MRI, Cryosection, and Segmentation data for the dataset after partial volumetric registration of MRI and Cryosection data. The train set comprises $10$,$000$ samples and the test set comprises $2$,$000$ samples. The source code for our approach will be released later for research purposes. 

\begin{figure*}[!h]
        \centering
        \setlength{\tabcolsep}{1pt} 
        \scriptsize
\begin{tabular}{lccccccccc}
& \multicolumn{3}{c}{$\overbrace{\rule{1.5in}{0pt}}^{\textrm{Sample 1}}$} & \multicolumn{3}{c}{$\overbrace{\rule{1.5in}{0pt}}^{\textrm{Sample 2}}$} & \multicolumn{3}{c}{$\overbrace{\rule{1.5in}{0pt}}^{\textrm{Sample 3}}$} \\
 & {256$\times$256} & {128$\times$128} & {64$\times$64} & {256$\times$256} & {128$\times$128} & {64$\times$64} & {256$\times$256} & {128$\times$128} & {64$\times$64}\\
\raisebox{0.0in}{\rotatebox{90}{GT Cryo.}} &
\includegraphics[width=\tileFraction\linewidth]{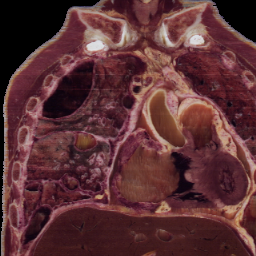} &
\includegraphics[width=\tileFraction\linewidth]{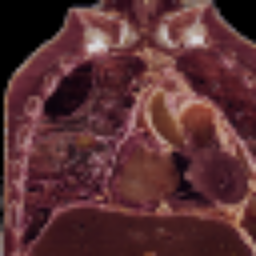}&
\includegraphics[width=\tileFraction\linewidth]{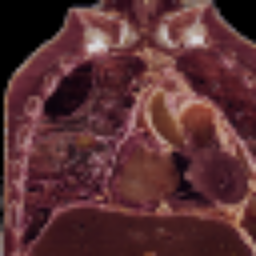}&
 \includegraphics[width=\tileFraction\linewidth]{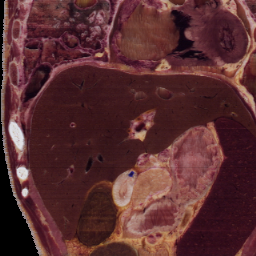}&
 \includegraphics[width=\tileFraction\linewidth]{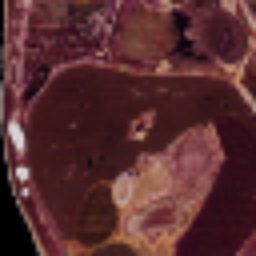}&
 \includegraphics[width=\tileFraction\linewidth]{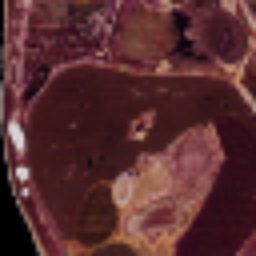}&
 \includegraphics[width=\tileFraction\linewidth]{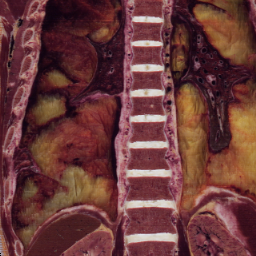}&
 \includegraphics[width=\tileFraction\linewidth]{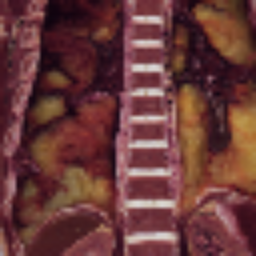}&
 \includegraphics[width=\tileFraction\linewidth]{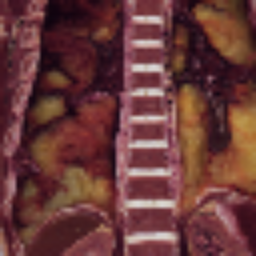}\\
\raisebox{0.05in}{\rotatebox{90}{Input MRI}} &
\includegraphics[width=\tileFraction\linewidth]{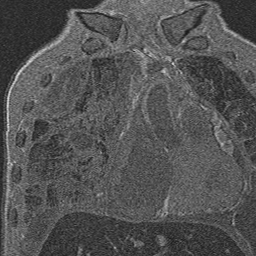}&  \includegraphics[width=\tileFraction\linewidth]{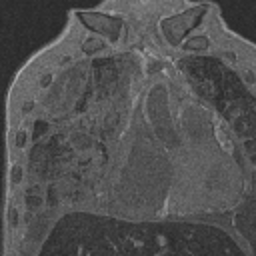}&
\includegraphics[width=\tileFraction\linewidth]{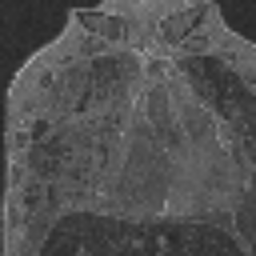}&
\includegraphics[width=\tileFraction\linewidth]{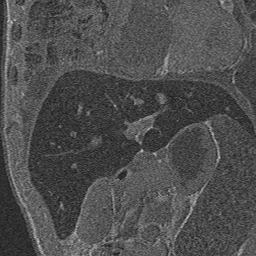}&
\includegraphics[width=\tileFraction\linewidth]{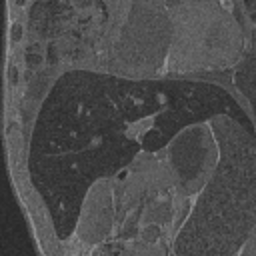}&
\includegraphics[width=\tileFraction\linewidth]{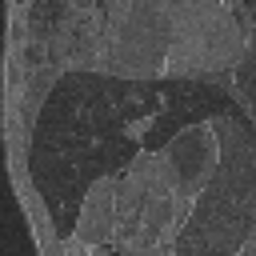}&
\includegraphics[width=\tileFraction\linewidth]{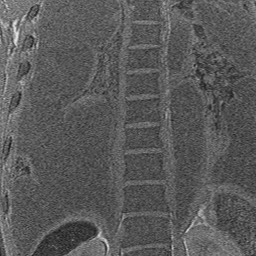}&
\includegraphics[width=\tileFraction\linewidth]{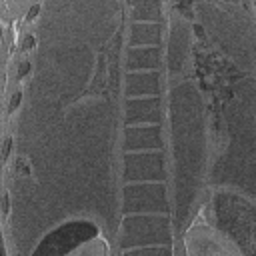}&
\includegraphics[width=\tileFraction\linewidth]{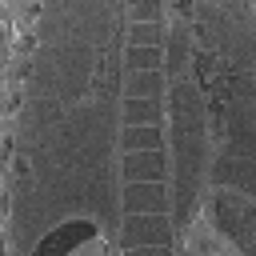}\\
\raisebox{0.15in}{\rotatebox{90}{Output}} &
\includegraphics[width=\tileFraction\linewidth]{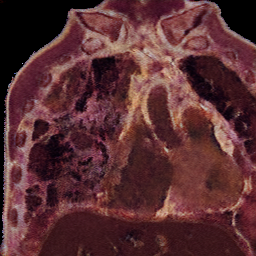}&
\includegraphics[width=\tileFraction\linewidth]{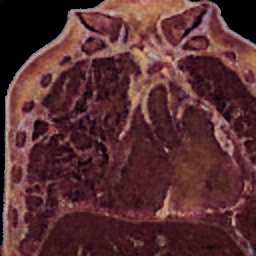} &
\includegraphics[width=\tileFraction\linewidth]{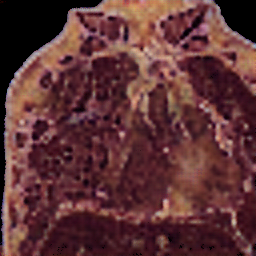} &
\includegraphics[width=\tileFraction\linewidth]{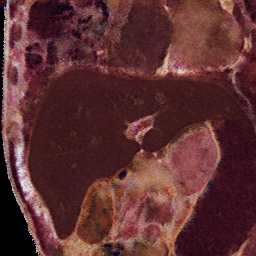}& 
 \includegraphics[width=\tileFraction\linewidth]{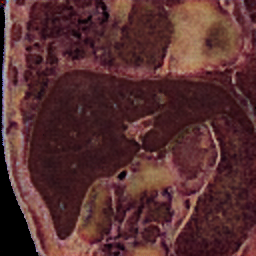}&
  \includegraphics[width=\tileFraction\linewidth]{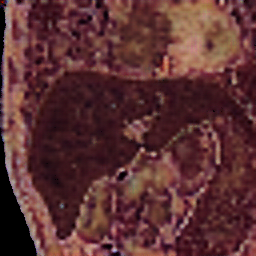}&
\includegraphics[width=\tileFraction\linewidth]{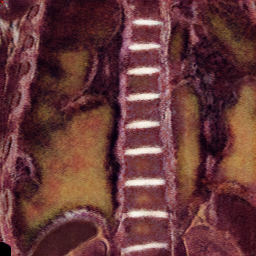}   & 
\includegraphics[width=\tileFraction\linewidth]{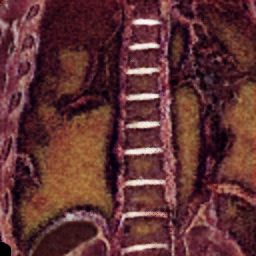}& 
\includegraphics[width=\tileFraction\linewidth]{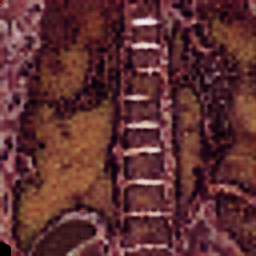} 
\end{tabular}%
\caption{MRI colorization results on three scales of resolution.}
\label{fig:results}
\end{figure*}

During inference the primary generator $\mathcal{G}_{m\rightarrow \hat{c}}$ takes input MRI image for colorization. While Cryosection and segmentation ground truth images are needed during the training of our model, these are not required at inference time. The results of our colorization are displayed in Fig.~\ref{fig:results} for four sample images taken across the human body. These results demonstrate the structural accuracy of our colorization in alignment with the input MRI, as well as the detailed and vivid textures of the ground truth Cryosection images. Since our approach utilizes organ-level segmentation information, it clearly demonstrates how different tissues, such as bones, muscles, cartilage, liver, and others, are accurately colorized. Furthermore, to demonstrate the scale-invariance of our approach, we present colorization results for input MRI images of varying sizes: $256 \times 256$, $128 \times 128$, and $64 \times 64$ in Fig.~\ref{fig:results}. We observe that as the resolution decreases, our method effectively handles the changes in structural information for colorization, maintaining minimal perceptual loss in the colorized outputs.

\subsection{Evaluation metrics}
We use image comparison metrics to evaluate our approach, utilizing evaluation metrics categorized broadly into structural similarity, color similarity, and textural similarity. We include the following evaluation metrics on the testing set in our comparisons: colorfulness (CF) score \cite{hasler2003measuring}, Structural Similarity Index (SSIM) \cite{wang2004image}, Multiscale Structural Similarity Index (MS-SSIM) \cite{wang2003multiscale}, Feature Similarity Index (FSIM) \cite{zhang2011fsim}, and Structural Texture Similarity Index (STSIM) \cite{zujovic2009structural}. Note that the generated colorized MRI images should be structurally similar to the input greyscale MRI and the colors in the output should be similar to Crysoction data. CF depicts the diversity and vividness of colorized MRI however, a high CF score does not always signify good visual quality. Therefore, we use \(\triangle\)CF \cite{Kang2023ICCV} which computes the difference between the CF scores of the Cryosection image and the generated one. SSIM is used to measure the structural consistency between the colorized MRI and the input MRI. Multiscale Structural Similarity Index (MS-SSIM) \cite{wang2003multiscale} helps in capturing different levels of image details to capture human visual perception. SSIM is computed at varying scales of the image and these multiscale SSIM values are combined to obtain the MS-SSIM index. Feature Similarity Index (FSIM) \cite{zhang2011fsim} is an image assessment metric used to evaluate the perceptual similarity between ground truth Cryosection and colorized MRI. It focuses on edges, textures, and gradients for correlating perceptual information with human visual perception. Structural Texture Similarity Index (STSIM) \cite{4711975} extends the functionality of SSIM by incorporating textural information of the image. It evaluates the periodicity, directionality, and granularity of the image textures between the ground truth Cryosection and the colorized MRI. We believe that these metrics comprehensively cover the color and textural similarity between the colorized MRI, input MRI, and ground truth Cryosection images.

\subsection{Comparisons with state-of-the-art}
 We compare the results of our method with the state-of-the-art colorization methods for which we selected five competing colorization methods: ChromaGAN \cite{vitoria2020chromagan}, ColorFormer \cite{ji2022colorformer}, DDColor \cite{Kang2023ICCV},  APS \cite{9761546}, and ALDM \cite{Kim2024WACV}. All the comparisons are conducted by training these methods on the Visible Korean Human dataset.
 \begin{table*}[!htp]
\scriptsize
\centering
\renewcommand{\arraystretch}{1} 
\setlength{\tabcolsep}{6pt} 
\caption{Comparisons against state-of-the-art methods across different metrics.}
 \label{tbl:sota_comparisons}
    \begin{tabular}{lrrrrrr}\toprule
    \textbf{Methods}  & \textbf{CF} \(\uparrow\) & \textbf{\(\triangle\)CF} \(\downarrow\) & \textbf{SSIM} \(\uparrow\) & \textbf{MS-SSIM} \(\uparrow\) & \textbf{STSIM} \(\uparrow\) & \textbf{FSIM} \(\uparrow\) \\ \midrule
    \textbf{ChromaGAN \cite{vitoria2020chromagan}}  & 0.880 $\pm$ 0.012  & 0.033 $\pm$ 0.054 & 0.295 $\pm$ 0.192 & 0.620 $\pm$ 0.226 & 0.512 $\pm$ 0.286 & 0.736 $\pm$  0.064 \\ 
    \textbf{ColorFormer \cite{ji2022colorformer}}  & 0.858 $\pm$ 0.011 & 0.055 $\pm$ 0.055  & 0.108 $\pm$ 0.094 & 0.368 $\pm$ 0.111 & 0.484 $\pm$ 0.268 & 0.616 $\pm$ 0.177 \\
    \textbf{DDColor \cite{Kang2023ICCV}}  & 0.855 $\pm$  0.031 & 0.058 $\pm$ 0.035 &  0.441 $\pm$ 0.228 & 0.670 $\pm$ 0.200 & 0.561 $\pm$ 0.383 & 0.735 $\pm$ 0.148 \\ 
    \textbf{APS \cite{9761546}}  & 0.905 $\pm$  0.068 & 0.009 $\pm$ 0.002 &  0.687 $\pm$ 0.261 & 0.724 $\pm$ 0.212 & 0.870 $\pm$ 0.134 & 0.732 $\pm$ 0.125  \\ 
    \textbf{ALDM \cite{Kim2024WACV}}  & 0.831 $\pm$ 0.827 & 0.086 $\pm$ 0.760 & 0.364 $\pm$ 0.381 & 0.709 $\pm$ 0.632 & 0.722 $\pm$ 0.145 & 0.733 $\pm$ 0.125 \\
    \textbf{Ours}  & \textbf{0.908} $\pm$ \textbf{0.067}  & \textbf{0.005} $\pm$ \textbf{0.001} & \textbf{0.863} $\pm$ \textbf{0.115} &\textbf{0.866} $\pm$ \textbf{0.188} & \textbf{0.901} $\pm$ \textbf{0.039} & \textbf{0.755} $\pm$ \textbf{0.112}\\ \bottomrule    
    \end{tabular}
\end{table*}

\begin{figure*}[!htp]
        \centering 
        \setlength{\tabcolsep}{1pt} 
        \scriptsize
        \begin{tabular}{ccccccccc}
              \raisebox{0.22in}{\rotatebox[origin=c]{90}{Sample 1}} &
              \includegraphics[width=\tileFractionCom\textwidth]{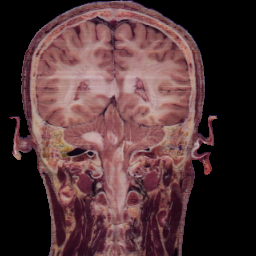} &
              \includegraphics[width=\tileFractionCom\textwidth]{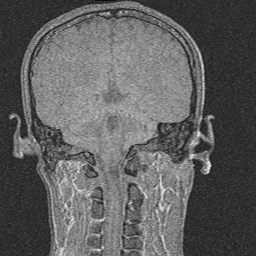}& 
              \includegraphics[width=\tileFractionCom\textwidth]{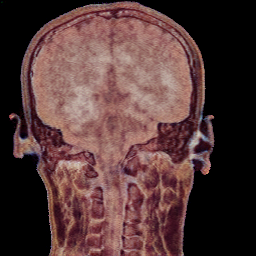} & 
              \includegraphics[width=\tileFractionCom\textwidth]{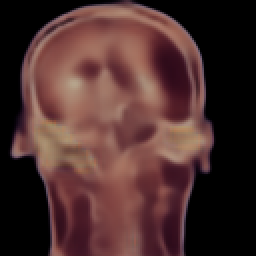}&
              \includegraphics[width=\tileFractionCom\textwidth]{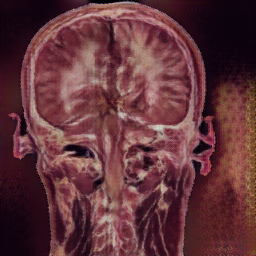}&
              \includegraphics[width=\tileFractionCom\textwidth]{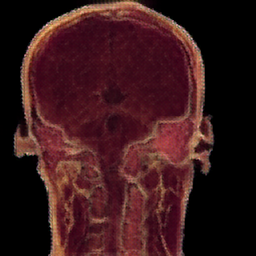}&
              \includegraphics[width=\tileFractionCom\textwidth]{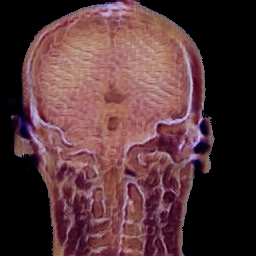}&
              \includegraphics[width=\tileFractionCom\textwidth]{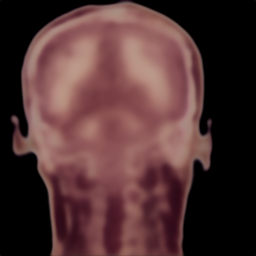}\\
              \raisebox{0.22in}{\rotatebox[origin=c]{90}{Sample 2}} &
              \includegraphics[width=\tileFractionCom\textwidth]{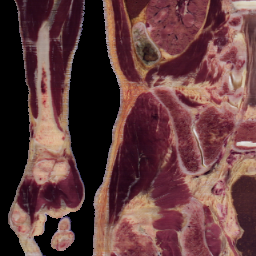} &
              \includegraphics[width=\tileFractionCom\textwidth]{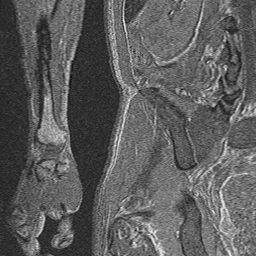}&  
              \includegraphics[width=\tileFractionCom\textwidth]{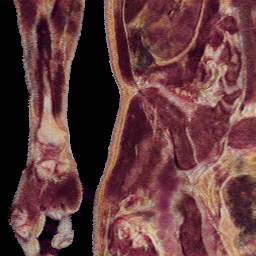} &
              \includegraphics[width=\tileFractionCom\textwidth]{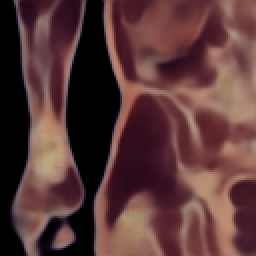}&
              \includegraphics[width=\tileFractionCom\textwidth]{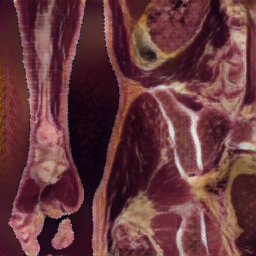}&
              \includegraphics[width=\tileFractionCom\textwidth]{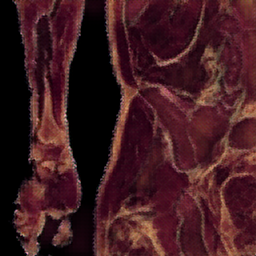}&
              \includegraphics[width=\tileFractionCom\textwidth]{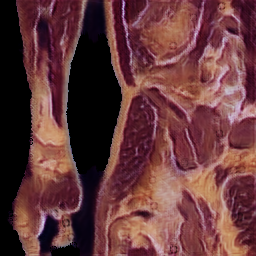}&
              \includegraphics[width=\tileFractionCom\textwidth]{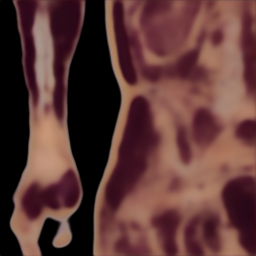}\\
              \raisebox{0.22in}{\rotatebox[origin=c]{90}{Sample 3}} &
              \includegraphics[width=\tileFractionCom\textwidth]{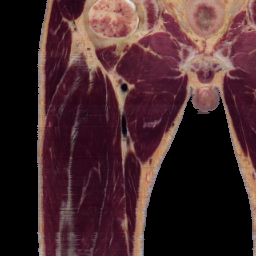} &
              \includegraphics[width=\tileFractionCom\textwidth]{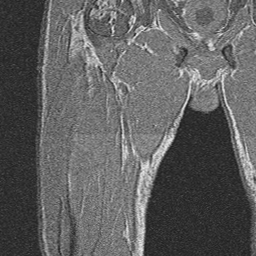}&
              \includegraphics[width=\tileFractionCom\textwidth]{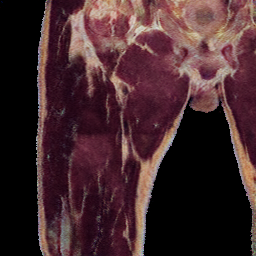}&
              \includegraphics[width=\tileFractionCom\textwidth]{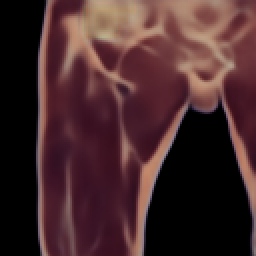}&
              \includegraphics[width=\tileFractionCom\textwidth]{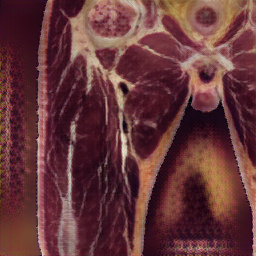}&
              \includegraphics[width=\tileFractionCom\textwidth]{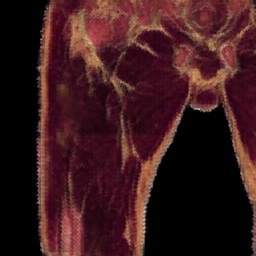}&
              \includegraphics[width=\tileFractionCom\textwidth]{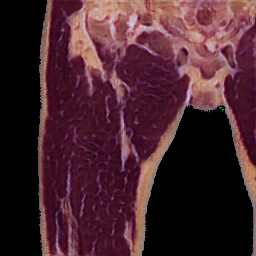}&
              \includegraphics[width=\tileFractionCom\textwidth]{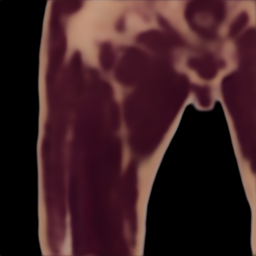}\\
              & GT Cryo. & Input MRI & Ours& Chroma- & Color- & DDColor \cite{Kang2023ICCV} & APS \cite{9761546} & ALDM \cite{Kim2024WACV} \\
              & & & & GAN \cite{vitoria2020chromagan} & Former \cite{ji2022colorformer} & &   &  
\end{tabular}
\caption{Quantitative comparison of MRI colorization with different methods.}
\label{fig:comparisons}
\end{figure*}

\begin{figure*}[!h]
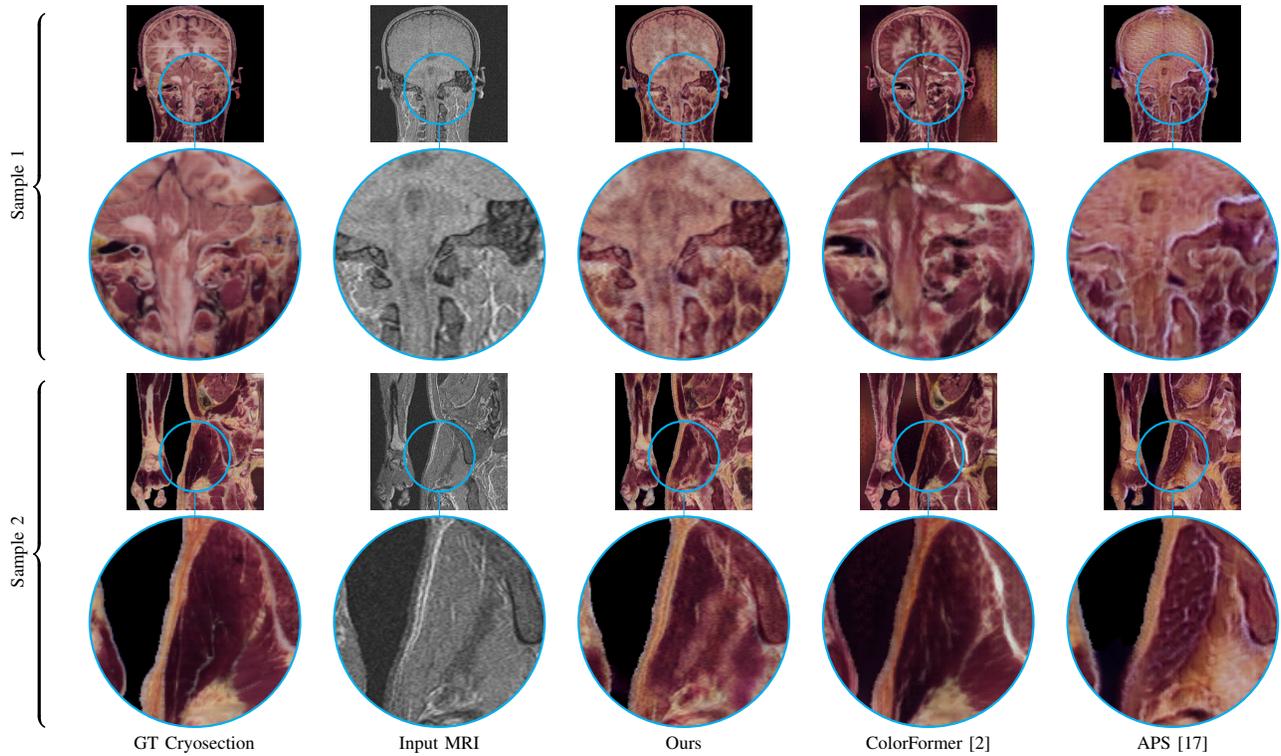

\centering
\scriptsize
\begin{tabular}{cccccc}
\raisebox{0.9in}{\rotatebox[origin=c]{90}{Sample 1}}
\begin{tikzpicture}[spy using outlines={circle,magnification=3,size=2.8cm,connect spies,every spy on node/.append style={thick}}]
    \draw [very thick,decorate,decoration={calligraphic brace,amplitude=4pt,mirror}] (-2,0.8) -- (-2,-3.8);
    \node {\includegraphics[width=\tileFraction\textwidth]{InputImages/cryo/sagittal_0.png}};
    \spy[color=cyan] on (0.0, -0.2) in node [below] at (0.0, -1);   
\end{tikzpicture} &
\begin{tikzpicture}[spy using outlines={circle,magnification=3,size=2.8cm,connect spies, line width=0.02mm, every spy on node/.append style={thick}}]
    \node {\includegraphics[width=\tileFraction\textwidth]{InputImages/mri/sagittal_0.png}};
    \spy[color=cyan] on (0.0, -0.2) in node [below] at (0.0, -1);    
\end{tikzpicture} &
\begin{tikzpicture}[spy using outlines={circle,magnification=3,size=2.8cm,connect spies,every spy on node/.append style={thick}}]
    \node {\includegraphics[width=\tileFraction\textwidth]{OurResults/sagittal_0.png}};
    \spy[color=cyan] on (0.0, -0.2) in node [below] at (0.0, -1);  
\end{tikzpicture} &
\begin{tikzpicture}[spy using outlines={circle,magnification=3,size=2.8cm,connect spies, line width=0.02mm, every spy on node/.append style={thick}}]
    \node {\includegraphics[width=\tileFraction\textwidth]{Comparison_ColorFormer/sagittal_0.png}};
    \spy[color=cyan] on (0.0, -0.2) in node [below] at (0.0, -1);    
\end{tikzpicture} &
\begin{tikzpicture}[spy using outlines={circle,magnification=3,size=2.8cm,connect spies,every spy on node/.append style={thick}}]
    \node {\includegraphics[width=\tileFraction\textwidth]{Comparison_aps/sagittal_0.png}};
    \spy[color=cyan] on (0.0, -0.2) in node [below] at (0.0, -1);   
\end{tikzpicture}\\
\raisebox{0.9in}{\rotatebox[origin=c]{90}{Sample 2}}
\begin{tikzpicture}[spy using outlines={circle,magnification=3,size=2.8cm,connect spies,every spy on node/.append style={thick}}]
    \draw [very thick,decorate,decoration={calligraphic brace,amplitude=4pt,mirror}] (-2,0.8) -- (-2,-3.8);
    \node {\includegraphics[width=\tileFraction\textwidth]{InputImages/cryo/sagittal_3.png}};
    \spy[color=cyan] on (0.0, -0.2) in node [below] at (0.0, -1);   
\end{tikzpicture} &
\begin{tikzpicture}[spy using outlines={circle,magnification=3,size=2.8cm,connect spies, line width=0.02mm, every spy on node/.append style={thick}}]
    \node {\includegraphics[width=\tileFraction\textwidth]{InputImages/mri/sagittal_3.png}};
    \spy[color=cyan] on (0.0, -0.2) in node [below] at (0.0, -1);    
\end{tikzpicture} &
\begin{tikzpicture}[spy using outlines={circle,magnification=3,size=2.8cm,connect spies,every spy on node/.append style={thick}}]
    \node {\includegraphics[width=\tileFraction\textwidth]{OurResults/sagittal_3.png}};
    \spy[color=cyan] on (0.0, -0.2) in node [below] at (0.0, -1);    
\end{tikzpicture} &
\begin{tikzpicture}[spy using outlines={circle,magnification=3,size=2.8cm,connect spies, line width=0.02mm, every spy on node/.append style={thick}}]
    \node {\includegraphics[width=\tileFraction\textwidth]{Comparison_ColorFormer/sagittal_3.png}};
    \spy[color=cyan] on (0.0, -0.2) in node [below] at (0.0, -1);   
\end{tikzpicture} &
\begin{tikzpicture}[spy using outlines={circle,magnification=3,size=2.8cm,connect spies,every spy on node/.append style={thick}}]
    \node {\includegraphics[width=\tileFraction\textwidth]{Comparison_aps/sagittal_3.png}};
    \spy[color=cyan] on (0.0, -0.2) in node [below] at (0.0, -1);   
\end{tikzpicture}\\
\hspace{0.4in}GT Cryosection & Input MRI & Ours & ColorFormer \cite{ji2022colorformer} & APS \cite{9761546}  
\end{tabular}
\caption{Zoom-in comparison with competing methods: ColorFormer \cite{ji2022colorformer} and APS \cite{9761546} produce appealing results, however a closer examination reveals structural dissimilarities with the input MRI and inconsistent colors compared to the ground truth Cryosection.}
\label{fig:zoomed_comparisons}
\end{figure*}

\begin{table*}[!h]
\scriptsize
\centering
\renewcommand{\arraystretch}{1} 
\setlength{\tabcolsep}{6pt} 
\caption{Quantitative measures of effectiveness of various components of our architecture.}
\centering
    \begin{tabular}{lrrrrrr}\toprule
    \textbf{Methods}  & \textbf{CF} \(\uparrow\) & \textbf{\(\triangle\)CF} \(\downarrow\) & \textbf{SSIM} \(\uparrow\) & \textbf{MS-SSIM} \(\uparrow\) & \textbf{STSIM} \(\uparrow\) & \textbf{FSIM} \(\uparrow\) \\ \midrule
    \textbf{Ours}  & 0.908 $\pm$ 0.067  & \textbf{0.005} $\pm$ \textbf{0.001}  & \textbf{0.863} $\pm$ \textbf{0.115} &\textbf{0.866} $\pm$\textbf{0.188}  & \textbf{0.901} $\pm$ \textbf{0.039} & \textbf{0.755} $\pm$ \textbf{0.112}\\    
    \textbf{A}$_1$: -cycle  & \textbf{0.914} $\pm$  \textbf{0.064} & -0.001 $\pm$ 0.067 &   0.640 $\pm$ 0.320 & 0.679 $\pm$ 0.315 & 0.885 $\pm$ 0.090 & 0.709 $\pm$  0.099 \\
    \textbf{A}$_2$: -seg  & 0.817 $\pm$ 0.016 & 0.017 $\pm$ 0.049   & 0.380 $\pm$ 0.352  & 0.708 $\pm$ 0.248 & 0.460 $\pm$ 0.374 & 0.573 $\pm$ 0.117 \\
    \textbf{A}$_3$: -seg, -pse. cryo  & 0.886 $\pm$ 0.072 & 0.027 $\pm$ 0.005 & 0.527 $\pm$ 0.384 & 0.137 $\pm$ 0.435 & 0.824 $\pm$ 0.130 & 0.628 $\pm$ 0.170\\
    \textbf{A}$_4$: +seg on rec. cryo  & 0.825 $\pm$  0.015 & 0.015 $\pm$ 0.051  & 0.388 $\pm$ 0.362 & 0.656 $\pm$ 0.256 & 0.503 $\pm$ 0.377 & 0.573 $\pm$ 0.120 \\
    \textbf{A}$_5$: -comp. activation  & 0.711 $\pm$ 0.111 & 0.203 $\pm$ 0.045    & 0.399 $\pm$ 0.353 & 0.559 $\pm$ 0.079 & 0.478 $\pm$ 0.386 & 0.541 $\pm$ 0.211 \\ \bottomrule    
    \end{tabular}
    \label{table:ablations}
\end{table*}

\begin{figure*}[!h]
        \centering
        \setlength{\tabcolsep}{1pt} 
        \scriptsize
        \begin{tabular}{ccccccccc}\\
              \raisebox{0.22in}{\rotatebox[origin=c]{90}{Sample 1}} &
              \includegraphics[width=\tileFractionAb\textwidth]{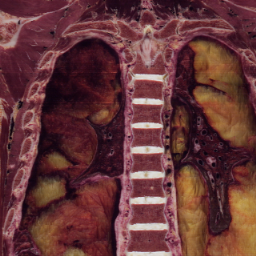}&
              \includegraphics[width=\tileFractionAb\textwidth]{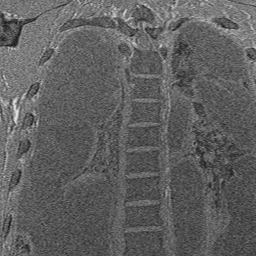}&      \includegraphics[width=\tileFractionAb\textwidth]{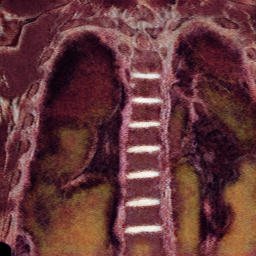}&    
              \includegraphics[width=\tileFractionAb\textwidth]{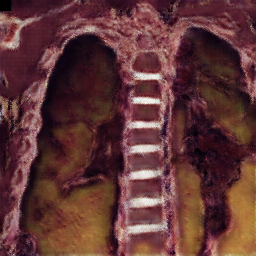}&
              \includegraphics[width=\tileFractionAb\textwidth]{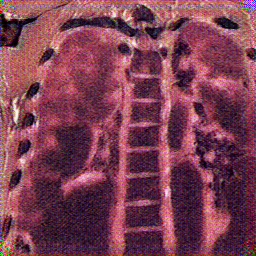}&              
              \includegraphics[width=\tileFractionAb\textwidth]{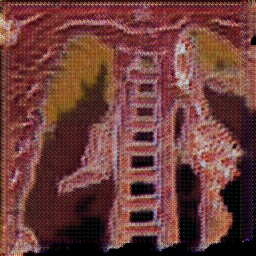}&
              \includegraphics[width=\tileFractionAb\textwidth]{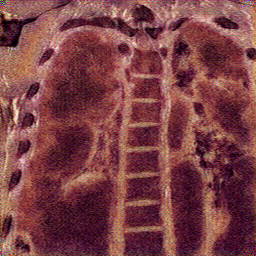}&
              \includegraphics[width=\tileFractionAb\textwidth]{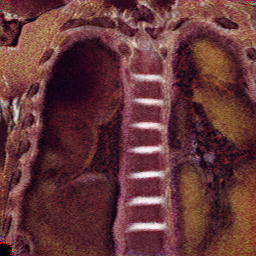}
              \\
              \raisebox{0.22in}{\rotatebox[origin=c]{90}{Sample 2}} &
              \includegraphics[width=\tileFractionAb\textwidth]{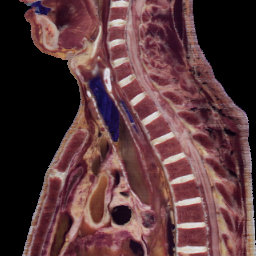} &
              \includegraphics[width=\tileFractionAb\textwidth]{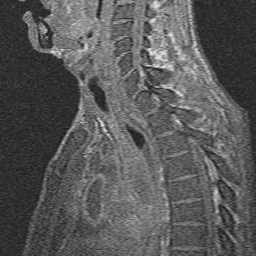}&  
              \includegraphics[width=\tileFractionAb\textwidth]{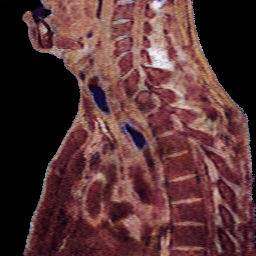} &
              \includegraphics[width=\tileFractionAb\textwidth]{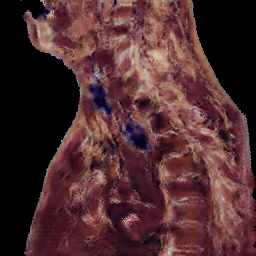}&
              \includegraphics[width=\tileFractionAb\textwidth]{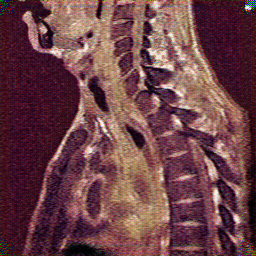}&              
              \includegraphics[width=\tileFractionAb\textwidth]{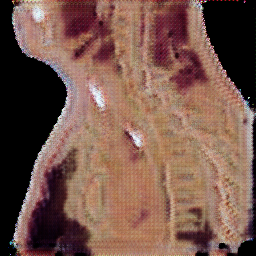}&
              \includegraphics[width=\tileFractionAb\textwidth]{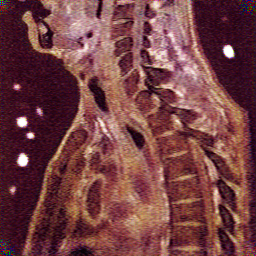}&
              \includegraphics[width=\tileFractionAb\textwidth]{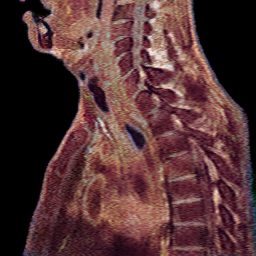}\\
              & GT Cryo. &  Input MRI & Ours & \textbf{A}$_1$ & \textbf{A}$_2$ & \textbf{A}$_3$ & \textbf{A}$_4$ &  \textbf{A}$_5$   \\
\end{tabular}
\caption{Qualitative comparison of the effectiveness of various components of our architecture. }
\label{fig:ablations}
\end{figure*}
\begin{figure*}[!h]
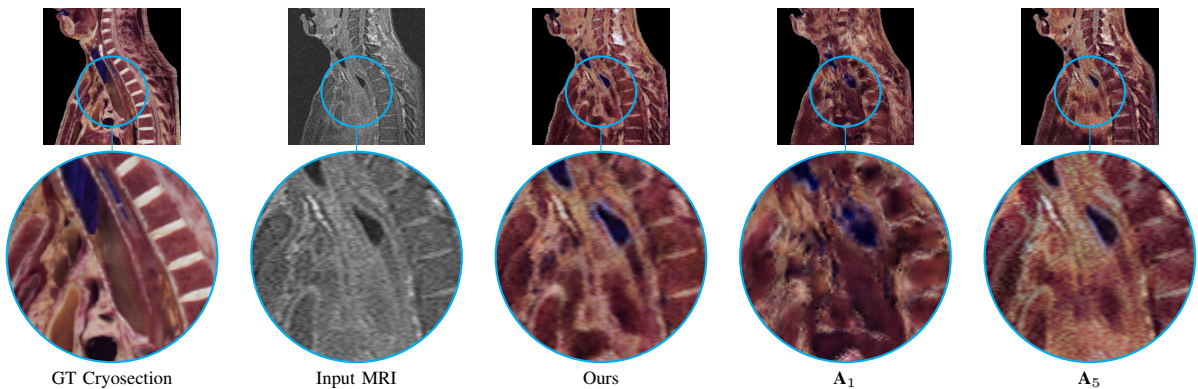

\centering
\scriptsize
\begin{tabular}{ccccc}
\begin{tikzpicture}[spy using outlines={circle,magnification=3,size=2.8cm,connect spies,every spy on node/.append style={thick}}]
    \node {\includegraphics[width=\tileFraction\textwidth]{InputImages/cryo/coronal_1.png}};
    \spy[color=cyan] on (0.0, -0.2) in node [below] at (0.0, -1);   
\end{tikzpicture} &
\begin{tikzpicture}[spy using outlines={circle,magnification=3,size=2.8cm,connect spies, line width=0.02mm, every spy on node/.append style={thick}}]
    \node {\includegraphics[width=\tileFraction\textwidth]{InputImages/mri/coronal_1.png}};
    \spy[color=cyan] on (0.0, -0.2) in node [below] at (0.0, -1);   
\end{tikzpicture} & 
\begin{tikzpicture}[spy using outlines={circle,magnification=3,size=2.8cm,connect spies,every spy on node/.append style={thick}}]
    \node {\includegraphics[width=\tileFraction\textwidth]{OurResults/coronal_1.png}};
    \spy[color=cyan] on (0.0, -0.2) in node [below] at (0.0, -1); 
\end{tikzpicture} &
\begin{tikzpicture}[spy using outlines={circle,magnification=3,size=2.8cm,connect spies, line width=0.02mm, every spy on node/.append style={thick}}]
    \node {\includegraphics[width=\tileFraction\textwidth]{ablation1/coronal_1.png}};
    \spy[color=cyan] on (0.0, -0.2) in node [below] at (0.0, -1);   
\end{tikzpicture} &
\begin{tikzpicture}[spy using outlines={circle,magnification=3,size=2.8cm,connect spies,every spy on node/.append style={thick}}]
    \node {\includegraphics[width=\tileFraction\textwidth]{ablation_cycleFullWithoutSE/coronal_1.png}};
    \spy[color=cyan] on (0.0, -0.2) in node [below] at (0.0, -1);   
\end{tikzpicture}\\
GT Cryosection & Input MRI & Ours & \textbf{A}$_1$ & \textbf{A}$_5$ 
\end{tabular}
\caption{Zoom-in comparison with competing ablation experiments: in \textbf{A}$_1$, removing cycle consistency degrades the geometric and textural quality of the result, while removing the compression-activation in \textbf{A}$_5$ results in erosion of fine structures and limited control over color.}
\label{fig:zoomed_ablation}
\end{figure*}

\paragraph{Quantitative comparisons}
We note that the colorization methods ChromaGAN, ColorFormer, and DDColor utilize the pre-trained VGG network and are originally trained on natural images. These methods lack knowledge of medical image distribution, so we retrain them on our medical dataset. In contrast, the two methods APS and ALDM are specifically designed to work with medical images and transform between CT and MRI images. We retrain these methods on our dataset to adapt them for MRI and Cryosection image modalities.

Table~\ref{tbl:sota_comparisons} shows a benchmark comparison of our method with previous colorization methods using various evaluation metrics. Our method exhibits the highest quality of the colorized MRI images in terms of maintaining structural integrity and texture/color synthesis. Our method achieves the lowest \(\triangle\)CF as compared to the previous colorization methods, demonstrating the color distribution similarity in colorized MRI and Cryosection data. ALDM incorporates a 3D diffusion model for cross-modality translation in the medical domain, however, it exhibits low similarity in the evaluation metrics for structures of MRI and colors of Cryosection data. APS performs better as compared to the other state-of-the-art methods however, our method also achieves the maximum SSIM and MS-SSIM score, indicating retention of structural similarity in the colorized MRI from the input MRI and maximum STSIM and FSIM scores for color similarity with the ground truth Cryosection. 

\paragraph{Qualitative comparisons} 
We compare the qualitative performance of our method with the previous colorization methods on the test set as shown in Fig.~\ref{fig:comparisons}.  The results generated by  ChromaGAN lack structures of the input MRI and have limited textural information corresponding to anatomical segmentation. ColorFormer generates additional structures that are absent in the input MRI and generates incorrect colors within the same organs. As shown in the close-up in Fig.~\ref{fig:zoomed_comparisons}, it introduces structures absent in the MRI in sample 1 generates bone colors within the muscle class in sample 2, and adapts structures of ground truth Cryosection instead of input MRI. Colorformer fails to capture the structural and chromatic correspondence of the multimodal data.  DDColor lacks color correspondence between colorized images and the Cryosection data and lacks color diversity among anatomical structures. Our method achieves superior results in structural similarity with the input MRI and distinct colors similar to the Crysosection data. As shown in the close-up in Fig.~\ref{fig:zoomed_comparisons} for sample 1, APS colorized the boundaries with colors absent in the Cryosection. It overlooked the colorization of structures of the input MRI in sample 2. However, results from our methods exhibit structural similarity with input MRI and corresponding colors of Cryosection. ALDM does not capture varying colors and textures for different organs in the body and lacks the structural information of the input MRI.

\subsection{Ablation study}
To evaluate the significance of different components within our proposed deep-learning model, we performed a comprehensive ablation study. The results of the ablation study are shown in Table~\ref{table:ablations} and Fig.~\ref{fig:ablations}.

\textbf{\textbf{A}$_1$: Without cycle consistency}
In this experiment, MRI colorization is achieved by discarding $\mathcal{L}_{rec}$ from $\mathcal{L}_{cyc}$ and optimizing only the forward pass of the cyclic generative adversarial network (GAN) with segmentation semantics. As shown in Fig.~\ref{fig:ablations}, this experiment led to distortions in the internal structures of the colorized MRI. In our architecture, the cyclic modality adaptation technique leverages the information of the source image reconstruction, aiding the structural consistency across multimodal unregistered data, as shown in the Fig.~\ref{fig:zoomed_ablation}. 

\textbf{\textbf{A}$_2$: Without segmentation network}
For this experiment, the semantic segmentation module $\mathcal{S}_{c\rightarrow \hat{s}}$ and segmentation loss $\mathcal{L}_{seg}$ are removed from our architecture, and only pseudo Cryosection synthesis executes cross-modality semantic information fusion for colorization. The absence of segmentation information leads to degradation in organ color variations in the colorized MRI as seen in the visual comparisons in Fig.~\ref{fig:ablations}. 

\textbf{\textbf{A}$_3$: Without segmentation network and pseudo Cryosection}
In this variant, we discard the semantic segmentation module along with the decoder $\mathcal{F}_{x\rightarrow c^\prime}$ that performs pseudo Cryosection synthesis. Consequently, the $\mathcal{L}_{ssim}$ loss is computed from the input MRI and the colorized MRI. As shown in Fig.~\ref{fig:ablations}, the colorized MRI is devoid of structural and color correspondence. 

\textbf{\textbf{A}$_4$: Segmentation on reconstructed Cryosection}
Here, we remove the pseudo Cryosection synthesis module $\mathcal{F}_{x\rightarrow c^\prime}$ and propagate the reconstructed Cryosection synthesized from the cyclic generative network to the segmentation network. Pseudo Cryosection is used for building semantic associations between input MRI and ground truth Cryosection that are only partially registered. As seen in Fig.~\ref{fig:ablations}, this results in degradation in the color distribution of the colorized MRI upon removing $\mathcal{F}_{x\rightarrow c^\prime}$.

\textbf{\textbf{A}$_5$: Without compression-activation mechanism}
In this variant, the compression-activation blocks in the skip connections are replaced with conventional convolutional layers to show the effectiveness and performance of our architecture comprising of compression-activation mechanism. The colorized MRI synthesized using this variant has limited color variations for anatomical regions and region-based global information since compression-activation mechanism facilitates noise suppression. As shown in the Fig.~\ref{fig:zoomed_ablation}, finer structures in the colorized MRI are absent.

\section{Discussion and challenges}
Our proposed colorization architecture performs realistic color transfer from the ground truth Cryosection data to an input MRI. We trained our architecture on MRI and Cryosection data with partial registration between the two modalities. Our system is capable of building a semantic association between the above two modalities via the segmentation of ground truth Cryosection. Comparison with the state-of-the-art methods shows superior performance of our method which can preserve MRI structures and synthesize colors and textures of ground truth Cryosection. Currently, our current approach is memory intensive since cross-modality networks require two pairs of generators and discriminators for the respective adapting distributions along with the segmentation network of corresponding modalities. This is not a limitation, however, it requires a high-end computing resource for training and it would be beneficial to reduce this memory constraint. The colorized MRI from our method inherits the noise and lack of high-frequency details from the input MRI. Our future work is to incorporate a method for improving the resolution of the input MRI data to match with the high details of the ground truth Cryosection. Further, the segmentation data used in our method corresponds to the Cryosection data only. As a future work, we would like to modify the presented architecture to segment MRI data as an unsupervised task based on the segmentation data of Cryosection.

\section{Conclusion}
The primary contribution of this work is the development of a generator architecture for synthesizing colorized MRI images with Cryosection-like textural details of various anatomical structures. Our cross-modality adaptation-based deep learning architecture is multiscale, and captures global structural details. By utilizing segmentation information, our model effectively adapts the color distribution of the ground truth Cryosection data, ensuring coherence between the anatomical structures of the input MRI and the resulting colorized images. Our approach demonstrates superior performance in both colorization and texture synthesis compared to existing methods.

\section{Compliance With Ethical Standards}
This research study was conducted using the Visible Korean Human data made available by KISTI via an agreement. 

\section{Acknowledgments}
This research was supported by the Science and Engineering Research Board (SERB) of the Department of Science and Technology (DST) of India (Grant No. CRG/2020/005792). The Visible Korean Human dataset for our research was provided by  the Korea Institute of Science and Technology Information (KISTI), South Korea.

\bibliographystyle{IEEEtran}
\bibliography{MRIColorization}
 
\end{document}